\documentclass{aa} %
\usepackage{epsfig}
\newcommand{\gsim}{\raisebox{-0.3ex}{\mbox{$\stackrel{>}{_\sim} \,$}}}

\begin{document}

   \thesaurus{08     
	      (08.16.6)}  
   \title{Revisiting the shape of pulsar beams}


   \author{D. Mitra and A. A. Deshpande}

   \offprints{dmitra@rri.ernet.in}

   \institute{Raman Research Institute, 
	      C. V. Raman Avenue, Bangalore - 560080, India\\
	      email:dmitra@rri.ernet.in, desh@rri.ernet.in} 

   \date{Received ? / Accepted ?}

   \maketitle

\begin{abstract}
 Characterizing the shape and evolution of 
pulsar radio emission beams
is important for understanding the observed 
emission. The various attempts by earlier 
workers investigating beam shapes
have resulted in widely different conclusions.
Using a carefully selected subset of the recently 
published multifrequency polarimetry 
observations of 300 radio pulsars 
(Gould \& Lyne, 1998),
we attempt to model the shape of pulsar beams. 
Assuming that the beam shape is elliptical, 
in general, and that it may depend on the 
angle between the rotation and the magnetic axes, we seek a 
consistent model where we also solve for
the dependence of the beam size on frequency. 
From the  six-frequency data on conal triple and multiple 
component profiles, we show that  
a) the pulsar emission beams follow a {\it nested 
cone} structure with at least 
{\em three distinct cones}, although only one or more 
of the cones may be active in a given pulsar; 
b) each emission cone is illuminated in the form of an 
annular ring of width typically about 20\% of the cone 
radius. 

Although some slight preference is evident for a model where
the beam is circular for an 
aligned rotator \& latitudinally compressed for an orthogonal 
rotator, the possibility that the 
beam shape is circular at all inclinations is found to be equally
consistent with the data. 
While the overall size scales as $P^{-0.5}$ (where 
$P$ is the pulsar period) as expected from the
notion of dipolar open field lines, we see no evidence
in support of the beam shape evolution with pulsar period.
\\
\\
\end{abstract}

\section{Introduction}Most widely accepted 
emission models assume that pulsar radiation is emitted
over a (hollow) cone centered around the magnetic dipole axis. 
The observed emission is generally 
highly linearly polarized with a systematic rotation of the 
position angle across the pulse profile. 
This behaviour, following Radhakrishnan
\& Cooke (1969), is interpreted in terms of the radiation being 
along the cone of the dipolar open field-lines 
emerging from the polar cap, and the plane of the 
linear polarization is that 
containing the field line associated with the emission 
received at a given instant.
During each rotation of the star, the emission beam 
crosses the observers line-of-sight resulting in a pulse 
of emission. The observed pulse profile thus corresponds 
to a {\it thin} cut across the beam at a fixed rotational
latitude. 
The information on the beam shape as a function of latitude, although 
generally not measurable directly, may 
be forthcoming from observations at 
widely separated frequencies, as 
emission at different frequencies is believed to 
originate at different heights from the star leading to changes
in beam size.
For this, the dependence of the radiation frequency on the height, 
the so called {\it radius-to-frequency mapping}, should be known a priori. 
Alternatively, it is possible to use the data on 
an ensemble of pulsars sampling a range of impact parameters.
However, it is important that all the pulsars in the sample 
form a homogeneous set in terms of the profile types 
etc. Several attempts to model the pulsar beam have used the 
latter approach.  Based on their study, 
Narayan and Vivekanand (1983) concluded that the beam is 
elongated in the latitude. Lyne \& Manchester 
(1988), on the other hand, have argued that the
beam is essentially circular 
(see also Gil \& Han 1996, Arendt \& Eilek 1998).
Based on the dipole geometry 
of the cone of open field-lines, Biggs (1990) found 
that the beam shape is a function of the
angle ($\alpha$) between the rotation and the magnetic axes. 
The reasons that all these analyses
predict different results could be manifold. For 
example, Narayan \& Vivekanand used a data set consisting 
of only 16 pulsars and assessed the beam axial ratio on the basis 
of the total change in the position angle of the 
linear polarization across the pulse profile. 
Apart from poor statistics, their analysis suffered 
from the large uncertainties in the polarization measurements available 
then. Lyne \& Manchester (1988) used a much larger data set in 
comparison and examined the distribution of normalized impact parameter 
$\beta_{n}\equiv\beta_{90}/\rho_{90}$, where $\beta_{90}$ \& 
$\rho_{90}$ are the impact angle and the beam radius computed for
$\alpha = 90^{\circ}$. 
Based on their observation 
that the distribution of $\beta_{n}$ is `essentially uniform', 
they concluded that the beams are circular in shape.  
The apparent deficit at large $\beta_{n}$
is attributed to a luminosity bias. 
It is worth noting 
that the deficit is seen despite the
fact that $\beta_{n}$ overestimates the {\it true} $\beta/\rho$ 
(given that they disregarded the sign of 
$\beta$), this is particularly so at large $\beta$ values.

Biggs (1990) used the same data set as well as the $\beta_{n}$ 
distribution as used by Lyne and Manchester (1988),
but drew attention to a `peak' in the distribution at low $\beta_{n}$. 
The shapes of the polar cap defined by the region of 
open field lines, as derived by Biggs, show that 
the beam is circular for an aligned rotator, but undergoes 
compression along the latitudinal direction 
with increasing inclination $\alpha$.
     
    In this paper, we address this question within the basic 
framework advanced by Rankin (1993a) which, at the least, 
is qualitatively different from that of Lyne \& Manchester (1988). 
The classification scheme (Rankin, 1983a), based on the phenomenology 
of pulse profiles, polarization and other fluctuation properties 
etc., provides a sound basis for explicit distinction between the
core and the conal components, with each of them following a predictable 
geometry 
(see also Oster \& Sieber 1976; Gil \& Krawczyk 1996 for {\it conal beams}).
Lyne \& Manchester (1988), on the other hand, prefer 
to interpret the observed variety in pulse shape and other properties
as a result of patchy illumination, rather than any particular 
pattern within the radiation cone.
    The observed differences in the properties of pulse components 
are then to be understood as gradual changes as a function of the 
distance from the center of the basic emission cone. 
Their analysis thus naturally disregards the possible 
existance of conal features.

Assuming the possibly confined `conal-component' geometry and by 
accounting for all the relevant geometrical effects, we re-examine 
the shape of pulsar beams and their frequency dependence. 
Recently published multifrequency polarization data, at six 
frequencies in the range between 234-1642 MHz 
(Gould \& Lyne, 1998),
has made this investigation possible.

\section{Data set}
For the present investigation requiring reliable estimates of $\alpha$ 
\& $\beta$, we use the data set comprised of only those 
pulsars whose pulse profiles are identified as `triple' ($\bf{T}$) or
`multiple' ($\bf{M}$), as classified by Rankin (1993a, 1993b). 
The reason for the choice is that the $\bf{T}$ and $\bf{M}$ 
pulsars show a core component in addition to the
conal components, so that a reliable estimation of the angle 
($\alpha$) between the rotation axis and the magnetic axis
is possible, using Rankin's (1990) method.
In this method, the ratio of the observed core-width                  
to the limiting width ($2.45^{\circ} P^{-0.5}$)
is interpreted as the geometric factor 
$1/\sin(\alpha)$, providing by far the most reliable
estimates of $\alpha$. For the conal doubles and conal singles, 
devoid of any core component, the estimates of $\alpha$ are less reliable. 
The core singles are naturally excluded from this analysis of the
conal emission geometry.  For each pulsar in our selected sample,
we define the conal width as the separation between the peaks 
of the outermost conal components. It is important to note that 
the nominally `central' core component, which is argued to originate
closer to the stellar surface, may not necessarily be along the cone axis.
Such a possibility is clearly reflected in many pulse profiles 
where the core component is displaced from the `center'
definable from the conal
components. Hence, the location of the core component is 
disregarded in our estimation of the conal separation.
Columns 1 and 2 of table 1 list 
the name and profile type of these pulsars. Columns 3 to 8
list the calculated widths of the pulsars at frequencies 234, 
408, 610, 925, 1400, and 1642 MHz respectively. Column 9 gives 
the pulsar period in seconds. Columns 10 and 11 list the 
$\alpha$ and $\beta$ values of the pulsars taken from Rankin (1993b).

\begin{figure}
\epsfig{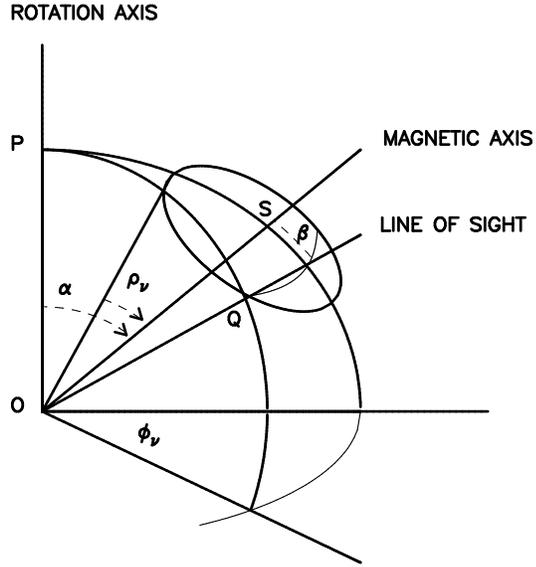}
\caption[]{Schematic representation showing the geometry of the 
pulsar emission region.}
\label{fig:fig1}
\end{figure}
 Rankin (1990) has estimated the inclination
angle $\alpha$ using the relation,
$\sin(\alpha) = 2.45^{\circ} P^{-0.5}/W_{\rm{core}}$, where $W_{\rm{core}}$ 
is the half-power width of the core component
(at a reference frequency 1 GHz)
and the period $P$ is in seconds. The impact angle $\beta$ has been
estimated based on the rotating vector model 
of Radhakrishnan \& Cooke (1969), using the relation 
$\sin(\beta) =(d\chi/d\phi)_{\rm{max}} /\sin(\alpha)$, where 
$(d\chi/d\phi)_{\rm{max}}$ is the maximum rate of change of
the polarization angle $\chi$ with respect to the
longitude $\phi$.

In the following analysis,
we treat the different frequency measurements on a 
given pulsar as `independent' inputs much the same way as
the data on different pulsars, since the pulsar beam size 
is expected to evolve with frequency. Thus, at different frequencies one 
obtains independent cuts (at different $\beta/\rho$) across 
the beam, though $\beta$ remains constant for a given pulsar. This 
increases the number of independent constraints by a usefully large 
factor. In fact, we would like to contrast this approach with the 
one where, for each pulsar, one obtains a best fit frequency 
dependence of the observed widths and then uses the data to 
obtain the width at a chosen reference frequency. The latter 
approach fails to take into account the dependence of the observed widths 
on $\beta/\rho$ that is inherent for any non-rectangular shape of the beam.

\begin{figure}
\epsfig{file=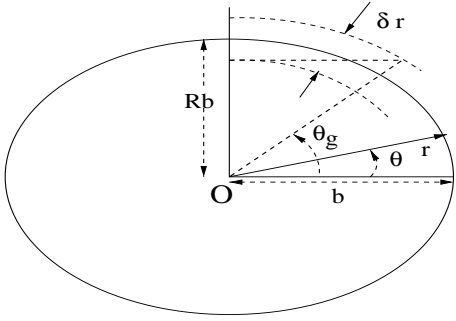,height=6cm,width=4.2cm,clip=true,angle=270}
\caption[]{ Schematic representation of an elliptic pulsar 
beam of axial ratio $R$ with the longitudinal and the 
latitudinal axis as $b$ and $Rb$ respectively. $\delta r$ is 
the width of the emission beam cone. 
See text for discussion on the
connection between the `gap-angle' $\theta_g$ and $\delta r/r$.} 
\label{fig:fig2}
\end{figure}
\section{A direct test for the shape of beams}
The Fig~\ref{fig:fig1}
is a schematic diagram illustrating the geometry of pulsar 
emission cone. The emission cone, with half-opening angle $\rho_{\nu}$,
sweeps across the observers line-of-sight with an impact parameter
(distance of closest approach to the magnetic axis) $\beta$. 
The spherical triangle PQS (refer to Fig.~\ref{fig:fig1}) 
relates the angles $\alpha$,
$\beta$ and the profile half-width $\phi_{\nu}$  
to the beam radius $\rho_{\nu}$
by the following relation
(Gil, Gronkowski \& Rudnicki 1984),
\begin{equation}
\sin^{2}(\rho_{\nu}/2)= 
\sin^{2}(\phi_{\nu}/2) \sin(\alpha) \sin(\alpha + \beta) + \sin^{2}(\beta/2)
\label{eq:modeq}
\end{equation}
The subscript $\nu$ in  $\rho_{\nu}$ and $\phi_{\nu}$ denotes that these
quantities depend on frequency $\nu$. This equation assumes that 
the cone is circular, in which case
$\rho_{\nu}$ becomes independent of $\beta$.
\begin{figure}
\epsfig{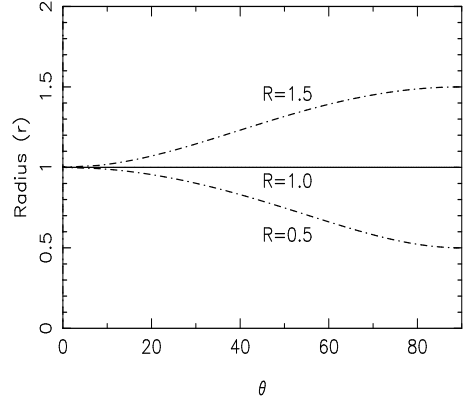}
\caption[]{The above curves illustrate the normalized 
variation of r with $\theta$
(refer to figure~\ref{fig:fig2}) with three different values of $R$.}
\label{fig:fig3}
\end{figure}

  In reality, the beam may not be circular, but rather elliptical
with, say, $R$ the axial ratio and b the longitudinal
semi-axis of the ellipse as shown
in Fig.~\ref{fig:fig2}. It is easy to see that 
the length of the radius vector r depends 
on the angle $\theta$ (with the longitudinal axis) when $R$ is not equal to 1.
The variation of r as a function of $\theta$ 
for three different $R$ values (namely 1, 1.5 and 0.5) are
shown as examples in Fig.~\ref{fig:fig3}. 
The $\rho_{\nu}$, determined assuming that the cone shape is circular
(as in Rankin 1993b) is indeed a measure of the
radius vector r, once the period and frequency dependences are corrected
for. Such data on (r,$\theta$) spanning a wide enough range in $\theta$ can
therefore be examined to seek a consistent value of the axial-ratio $R$. 
However, if $R$ is a function of $\alpha$, as suggested by Biggs (1990), then
the (r,$\theta$) samples would show a spread bounded by the curves 
corresponding to the maximum and minimum values of $R$. 

Such an examination of the present data suggests
a spread below the line for $R=1$, 
indicating that the beam
deviates from circularity and that the spread could be due to the 
$\alpha$ dependence of $R$. However, this 
deviation from circularity is not very significant. 
We discuss this in detail later in section 5.

  We have also examined the $\rho_{\nu}$ values obtained by Rankin (1993b) 
through such a test. However, no significant deviation
from circular beams was evident. We became aware 
of a similar study by C.-I. Bj$\ddot{o}$rnsson (1998), 
also with a
similar conclusion. We note that the 
only difference between our estimates of $\rho_{\nu}$ and those of Rankin
is in the definition of the conal widths. Rankin defines
the width as the distance between the outer half-power points (rather 
than the peaks) of the two conal outriders, and 
the widths were then
`interpolated' to a reference frequency of 1 GHz. Such estimates are 
prone to errors due to mode changes, differing 
component shapes etc., and to the effects of 
dispersion \& scattering (some of which she attempted to accommodate).
We measure the widths as the peak-to-peak separations 
of the outer conal components,
which are less sensitive to the sources of
error mentioned above. We have also confirmed 
(in the PSRs 0301+19, 0525+21, 0751+32, 1133+16, 1737+13, 2122+13
and 2210+29 using the data from Blaskiewicz et al. 1991)
that the `peaks' of the conal components are symmetrically 
placed with respect to the ``zero-longitude" (associated with
the maximum rate of change of the position angle), which is not
always true for the outer half-power points.
\begin{table*}
\begin{flushleft}
\caption[]{ The table lists the pulsar name and the widths measured at
6 different frequencies from the observations of 
Gould \& Lyne (1998).
In several cases the widths could not be estimated
due either to poor quality profiles or to absence of data.
The $\alpha$, $\beta$ values are taken from Rankin (1990, 1993b).
$LM$ indicates
that the $\beta$ value (for PSR 0656+14 and 1914+09) is taken from 
Lyne \& Manchester (1988).}
\begin{tabular}{lccccccccccccc }
\noalign{\smallskip}
\hline
{\bf Pulsar}&{\bf Profile}&\multicolumn{6}{c}{\bf Width in deg}&{\bf Period}&{\bf $\alpha$}& {\bf $\beta$}\\ 
\cline{3-8}
{\bf Bname}&{\bf Class}&{\bf $W_{234}$}&{\bf $W_{408}$}&{\bf $W_{610}$}&{\bf $W_{925}$}&{\bf $W_{1400}$}&
{\bf $W_{1642}$}&{\bf (sec)}&{\bf (deg)}& {\bf (deg)}\\ 
\noalign{\smallskip}
\hline
0329+54& \bf{T}& 25.4 &    23.3&        21.8 &     21.8&   21.2&     20.7& 0.714518&  30&  2.1\\
0450-18& \bf{T}& 16.6 &    14.5&        13.5 &     12.9&   12.4&     11.9& 0.548937&  24&  4\\
0450+55& \bf{T}& 27.3 &    20.7&        20.7 &     24.6&   22.0&     22.0& 0.340729&  32&  3.3\\
0656+14& \bf{T}& 27.9 &    21.7&        17.8 &     25.5&   20.1&     17.8& 0.384885&  30&  8.2 (LM)\\
0919+06& \bf{T}&18.1  &    16.5&        14.6&      11.5&   10&       8.4 & 0.430619&  48&  4.8     \\
~~~~~~~& ~~~~~~       &~~~~~   & ~~~~ ~~     & ~~~ ~~  & ~~ ~~~& ~~ ~~   & ~~~~~   &~~  & ~~~&  \\
1508+55& \bf{T}&  -   &    12.0&        8.57 &     11.6&   10.9&     10.5& 0.739681&  45&  -2.7\\
1541+09& \bf{T}& 126.5&   107.8&        105.4&    96.0&    91.4&     84.3& 0.748448&   5&  0.0 \\
1738-08& \bf{T}&-     &    14.6&        13.7&     13.6&    12.6&     12.1& 2.043082&  26&  1.7     \\
1818-04& \bf{T}&-     &    10.7&         8.2&     9.20&    8.8 &     8.5 & 0.598072&  65&  3.5     \\
1821+05& \bf{T}& 36.2 &    32.1&        29.4 &     29.4&   26.6&     26.6& 0.752906&  32&  1.7       \\
       &              &        &             &         &       &         &         &    &    & \\
1911+13& \bf{T}& -    &    12.3&        10.7&     12.0&    11.6&     11.0& 0.521472&  52&  1.9     \\
1914+09& \bf{T}& -    &    10.9&        12.6&      8.9&     8.5&     8.1 & 0.270254&  52&  7.3 (LM)     \\
1917+00& \bf{T}& -    &    8.3 &         8.1&      8.0&     7.2&      6.7& 1.272255&  81&  1.3     \\
1918+19& \bf{T}&-     &    49.1&        42.7&     41.3&    41.3&     38.7& 0.821034&  12&  -4.6     \\
1919+14& \bf{T}& -    &    22.3&        20.7&     18.7&    19.7&     17.1& 0.618179&  26& -6.4     \\
       &              &        &             &         &       &         &         &    &    & \\
1919+21& \bf{T}&-     &    7.17&         6.7&     8.2&     7.6&      7.4 & 1.337301&  45&  -3.7     \\
1920+21& \bf{T}& -    &    15.1&        10.1&     14.4&    14.0&     13.2& 1.077919&  44&  1.1     \\
1944+17& \bf{T}&-     &    25.2&        23.3&     33.0&    33.0&     31.0& 0.440618&  19&  6.1     \\
2045-16& \bf{T}& -    &    12.9&        12.3&     11.6&    11.0&     10.7& 1.961566&  36&  -1.1     \\
2111+46& \bf{T}& 69.8 &    63.3&        59.4&     55.6&    53.0&     49.1& 1.014684&   9&  1.4     \\
       &              &        &             &         &       &         &         &    &    & \\
2224+65& \bf{T}&39.9  &    35.0&        31.1&     31.1&    31.1&     31.1& 0.682537&  16&  3.4     \\
2319+60& \bf{T}&21.8  &    18.7&        17.1&     15.0&    13.5&     13.5& 2.256487&  18&  2.2     \\
1804-08& \bf{M/T}&-   &    28.5&        12.9 &    16.2&    15.5&     14.2& 0.163727&  63&  5.1 \\
1910+20& \bf{M/T}&-   &    12.8&        11.5&     11.2&    10.8&      -  & 2.232963&  29&  1.5     \\
1952+29& \bf{M/T}&-   &    22.7&        21.6&     22.2&    21.0&     19.2& 0.426676&  30&  -7.2     \\
       &              &        &             &         &       &         &         &    &    & \\
2020+28& \bf{M/T}&12.9&    10.9&        10.1&     10.1&    9.74&     9.3 & 0.343401&  72&  3.6     \\
0138+59& \bf{M}&25.8  &    20  &        23.2&     20.6&    18.7&     17.4& 1.222948&  20&  2.2     \\
0402+61& \bf{M}&14.2  &    14.6&        10.7&     10.3&    10&       9.6 & 0.594573&  83&  2.2     \\
0523+11& \bf{M}&-     &    12.4&        10.8&     12.0&    11.6&     10.8& 0.354437&  78&  5.9     \\
0621-04& \bf{M}&18.5  &    21.2&        18.4&     18.0&    17.5&       - & 1.039076&  32&  0.0     \\
       &              &        &             &         &       &         &         &    &    & \\
1039-19& \bf{M}&15.4  &    -   &        11.5&     10.7&    10&       9.6 & 1.386368&  31&  1.7     \\ 
1237+25& \bf{M}&10.0  &    10.3&        10.0&     9.3&     9.0&      10.0& 1.382449&  53&  0.0     \\
1737+13& \bf{M}&-     &    17.4&        17.0&     16.1&    15.2&     13.8& 0.803049&  41&  1.9      \\
1831-04& \bf{M}&95.3  &    97.6&        95.3&     96.2&    93.0&     93.0& 0.290106&  10&  2.0     \\
1857-26& \bf{M}&-     &    32.5&        29.4&     26.3&    25.5&     24.8& 0.612209&  25&  2.2     \\
       &              &        &             &         &       &         &         &    &    & \\
1905+39& \bf{M}&-     &    15.1&        13.7&     13.1&    12.6&     11.7& 1.235757&  33&  2.1     \\
2003-08& \bf{M}& 55.6 &    40.0&        38.7&     33.6&    32.3&     31.0& 0.580871&  13&  3.3      \\ 
\noalign{\smallskip}
\hline
\end{tabular}
\end{flushleft}
\label{table:tab1}
\end{table*}
\normalsize

\section{The model of the pulsar beam}
We model the pulsar beam shape as
elliptical in general and express it analytically as,
\begin{equation}
\frac{\sin^{2}(\phi_{\nu}/2) \sin(\alpha) \sin(\alpha + \beta)}{\sin^{2}(\rho_{\nu}/2)} + \frac{\sin^{2}(\beta/2)}{\sin^{2}(R \rho_{\nu}/2)}=1 
\label{eq:mod}
\end{equation}
\noindent
While $\alpha$, $\beta$ and $\phi_{\nu}$ can be estimated, directly 
or indirectly, from observations, $R$ and $\rho_{\nu}$ are the two
parameters which in turn define the beam shape
and size--- and the available data set of $\bf{T}$ 
and $\bf{M}$-profiles is expected to sample most 
of the $\mid \beta/\rho_{\nu}\mid$ range (0--1)
with reasonable uniformity.
The implicit assumption in this statistical approach is that a 
common description for $R$ \& $\rho_{\nu}$ is valid for all
pulsars. The common description should, however, 
account for relevant dependences on quantities, such as
frequency, period, $\alpha$, etc. properly.

\subsection{Frequency dependence of $\rho_{\nu}$}
The radio emission
at different frequencies is expected to originate at different altitudes
above the stellar surface, with  the higher 
frequency radiation 
associated with regions of lower altitude. This phenomenon  
known as {\em radius-to-frequency mapping}, finds 
overwhelming support from observations. Thorsett (1991)
has suggested an empirical relation for the observed pulse width  
as a function of frequency, which seems to 
provide adequate description of the observed behaviour.
We adopt a similar relation for
the frequency evolution of the beam radius $\rho_{\nu}$ as follows
\begin{equation}
\rho_{\nu} = \hat{\rho}(1 + K \nu^{-\zeta}),
\label{eq:rho}
\end{equation}  
\noindent
where $\hat{\rho}$ is the value of $\rho_{\nu}$ at infinite 
frequency, $\zeta$ the spectral index, 
and $K$ a constant. Note that both $\zeta$ \& $K$  
are expected to have positive values, so that  
the minimum value of $\rho_{\nu}$ is $\hat{\rho}$, which should 
correspond to the angular size of the polar cap.

\subsection{Period dependence on $\rho_{\nu}$}
Rankin (1993a) has demonstrated 
(see also Gil, Kijak \& Seiradakis 1993; Kramer et al. 1994)
that the beam radius $\hat{\rho}$ varies as 
$P^{-0.5}$ (where $P$ is the period of the pulsar), a result
which is in excellent agreement with that expected from a
dipole geometry (Gil 1981).
Eq~\ref{eq:rho} thus takes the form
\begin{equation}
\rho_{\nu} = \rho_{\circ} (1 + K \nu^{-\zeta}) P^{-0.5},
\label{eq:rho1}
\end{equation}  
\noindent 
where $\rho_{\circ}$ is the minimum beam radius for $P = 1$ sec.

\subsection{Functional dependence of $R$ on $\alpha$}
Biggs (1990) has suggested that $R$ should be a 
function of $\alpha$, such that  
the beam shape is circular for $\alpha=0$ and is 
increasingly compressed in the
latitudinal direction as $\alpha$ increases to $90^{\circ}$. 
We therefore model the functional dependence of $R$ on $\alpha$ as 
$R = R_{\circ}\tau$, where $R_{\circ}$ is the axial ratio
of the beam at $\alpha = 0$, and $\tau$ is a function of $\alpha$.
According to Biggs (1990), $R_{\circ} = 1$ and $\tau$ is
given by
\begin{equation}
\tau(\alpha)\;=\;1 - K_{1}\times10^{-4} \alpha - K_{2}\times10^{-5}\alpha^{2},
\label{eq:biggs}
\end{equation} 
\noindent
where $K_{1}$, $K_{2}$ are constants and $\alpha$ is in degrees. 
Biggs finds that $K_{1}$ and $K_{2}$ are 3.3 and 4.4, 
respectively. We, however, treat $K_{1,2}$ as free parameters
in our model.

\subsection{The number of hollow cones}Based on the study
of conal components, Rankin (1993a) has argued for two nested 
hollow cones of emission-- namely, the outer and the inner cone.
Assuming the beams to be circular in shape, opening half 
angles of the two cones at 1 GHz were found 
to be $4.3^{\circ}$ and $5.7^{\circ}$, 
respectively. 

 During our preliminary examination of the present sample, we noticed a
need to allow for three cones of emission. To incorporate this feature 
in our model, we introduce two ratios, $r1 < 1$ and $r2 > 1$,
to define the size scaling of the 
inner-most and the outer-most cone, respectively, 
with reference to a `middle' cone, for which the detailed shape is defined.

Using the model here defined, we need to solve for
$R_{\circ}$, $\zeta$, $K$, $\rho_{\circ}$, $K_{1}$,
$K_{2}$, $r1$ and $r2$ in this three-conal-ring model. The parameter set
thus represents an `average' description of the beam.

\begin{table*}
\caption[]{The best-fit model parameters for the shape of conal beams. 
The error bars correspond to a 1$\sigma$ uncertainty.}
\begin{flushleft}
\begin{tabular}{cccccccc}                 
\multicolumn{8}{c}{\bf Model parameters}\\ \hline 
$ R_{\circ}$ & $\rho_{\circ}$ (deg)& $K$ & $\zeta$ & $r1$ & $r2$ & $K_{1}$ ($\rm{deg^{-1}}$) & $K_{2}$ ($\rm{deg^{-2}}$)\\ 
\cline{1-8}
$0.91 \pm^{0.2}_{0.1}$ &${4.8\pm 0.3}$ & $66\pm 10$ & $1 \pm 0.1$ & $0.8\pm 0.03$ & $1.3\pm 0.03$ & ${7.2\pm 0.2}$ & ${4.4\pm 0.3}$   \\ \hline
\end{tabular}
\end{flushleft}
\end{table*}

\section{Results and Discussion}An optimized grid search was performed for
suitable ranges of the parameter values and in fine enough steps. 
For $\zeta$, the search range allowed for both +ve and -ve values.
By definition, $r_{1}\leq 1$ and $r_{2}\geq 1$.
The best fit was obtained by minimizing the standard deviation $\sigma_{\circ}$
defined by
\begin{equation}
\sigma_{\circ}\;=\;\sqrt{\frac{\sum_{i=1}^{n} D_{i}^{2}}{N_{dof}}} \times \frac{180^{\circ}}{\pi},
\label{sigma}
\end{equation}
\noindent
where $D_{i}$ is the deviation of the $i^{th}$ data 
point from the nearest conal ring in the model and $N_{dof}$ denotes 
the degrees of freedom.
\begin{figure}
\epsfig{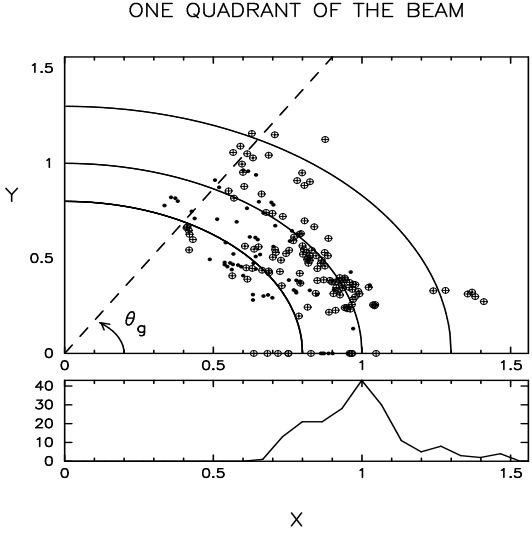}
\caption[]{Distribution of the (x,y) locations of the conal components
on a common scale. The three solid lines indicate the three emission cones
in the quadrant shown.The circles with crosses refers to pulsars with $\alpha$
values less than $45^{\circ}$ and the filled circles with $\alpha$ greater
than $45^{\circ}$.}
\label{fig:fig4}
\end{figure}
The factor $180/\pi$ gives $\sigma_{\circ}$ in units of
degrees under the small-angle approximation. 
Table 2 lists the parameter values which correspond to
the best fit for the entire sample set. 
With these values, the  eq.~\ref{eq:rho1} can now be rewritten as 
\begin{equation}
\rho_{\nu} = 4.8^{\circ} (1 + 66\; \nu^{-1}_{\rm MHz}) P^{-0.5},
\label{eq:rho2}
\end{equation}  
\noindent 
where $\rho_{\nu}$ is in degrees. This average description
for the `middle' cone applies also to the other two cones when 
$\rho_{\nu}$ is scaled by the ratio $r1=0.8$ or $r2=1.3$  
(for the inner and the outermost, respectively).
Fig~\ref{fig:fig4} shows the data (plotted to a common scale)
for one quadrant of the beam and the three solid curves
corresponding to the best fit cones. The points in the
figure, though corresponding to different pulsars and frequencies,
are translated to a common
reference scale appropriate for $P=1$ sec, 
$\alpha = 0$ and $\nu =\infty $. 

\begin{figure}
\epsfig{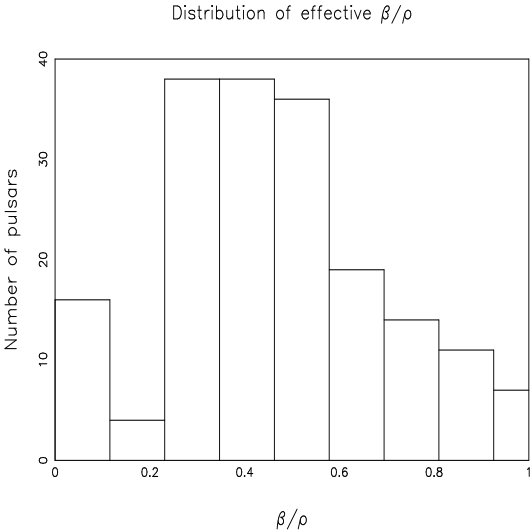}
\caption[]{Histogram of the distribution of effective $\frac{\beta}{\rho}$.}
\label{fig:fig5}
\end{figure}
We have assumed the period dependence of $\rho_{\nu}$ as  
$P^{-0.5}$, whereas Lyne and Manchester (1988) found a dependence 
of $P^{-\frac{1}{3}}$. We have examined the latter possibility 
and found that the difference in the standard deviation is at the 
level of 2.5-3 $\sigma$ and we cannot rule out the $P^{-\frac{1}{3}}$
law with confidence.
We have also checked for the dependence 
of $R$ on $\alpha$ by using 3 sub-sets, each of range $30^{\circ}$ 
in $\alpha$. 
The best fit values for $R$ in the different $\alpha$ segments 
are $1\pm^{0.4}_{0.2}$, $0.8\pm^{0.4}_{0.2}$ \& $0.5\pm^{0.4}_{0.2}$ for 
$\alpha$ ranges $0^{\circ}-30^{\circ}$, $30^{\circ}-60^{\circ}$ \&
$60^{\circ}-90^{\circ}$, respectively. This dependence of
$R_{\circ}$ on $\alpha$, even if it were significant, is quite
consistent with our values of $K1$, $K2$ (Table 2) as well as with
the results of Biggs (1990). However, given the uncertainties
in the $R$ estimates for the three ranges, it is not possible
presently to rule out a dependence of $R$ on $\alpha$.
Indeed, this part of the goodness-of-fit is negligible,
$\sigma_{\circ}$ (the standard deviation) is
$0.18^{\circ}$ when $K1$ and $K2 \neq 0$ and 
$0.2^{\circ}$ when $K1, K2 = 0$.  
Earlier Narayan \& Vivekanand (1983) had argued 
that $R$ is a function of the pulsar
period. To assess this claim, our sample was divided into three
period ranges and the corresponding $R$ estimates compared.
However, no period dependence was evident and
it was possible to rule out such a dependence
with high confidence. 

{\bf The number and thickness of conal rings:} 
As already noted and
can be seen in Figure~\ref{fig:fig4}, we do see evidence 
for a possible cone 
outside the two cones discussed by Rankin (1993a). 
Also, presence of a `further inner' cone          
has been suggested by Rankin \& Rathnasree (1997) 
in the case of PSR 1929+10. 
The pulsars suggestive of this outer cone (refer Figure~\ref{fig:fig4}) are
PSRs 0656+14, 1821+05, 1944+17 and 1952+29 (at frequencies 234 MHz and higher).
We have examined the possibility that these cases really belong to 
the central-cone, but are well outside of it due to an error in the 
assumed values of $\alpha$. We rule out the possibility as the implied 
error in $\alpha$ turns out to be too high to be likely.
It is important to point out that a noisy 
sample like the present one would appear increasingly
consistent, judging by the best-fit criterion, with models that 
include more cones. The question, therefore is whether 
we can constrain the number of cones by some independent 
method. In this context, we wish to discuss
{\it the noticeable deficit of points at high $\beta/\rho_{\circ}$}.
Since the deficit reflects the absence of conal singles
and conal doubles in our data set, the size of the related `gap' at
large $\theta$ values, can be used 
to estimate the possible thickness of the conal rings.
The absence of points at $\theta\gsim 60^{\circ}$ 
(Figure~\ref{fig:fig4}) suggests
that the conal rings are rather thin, since a radial thickness
$\delta r$ comparable to the ring radius would imply a wider gap in
$\theta$. To quantify this, we write the following relation,
\begin{equation}
\delta r= 2r \frac{(1-\sin\theta_{g})}{(1+\sin{\theta_{g}})}, 
\label{eq:deltar}
\end{equation}
\noindent
 where $\theta_{g}$ is the $\theta$ at the start of the gap
(as illustrated in Fig.~\ref{fig:fig2}).
With $\theta_{g} \sim 60^{\circ}$, $\delta r/r$ would be 
about 20\%.
The presence of more than one distinguishable
peak in the distribution of beam radii (shown in the bottom 
panel of Fig.~\ref{fig:fig4}) clearly indicates that 
the conal separation
is larger than the cone width. This combined with our cone-width
estimate suggests the number of cones is 3 (for the present range
of radii), providing an independent support for our model.
This picture is consistent with the estimates by Gil \& Krawczyk (1997)
and Gil \& Cheng (1999).

{\bf Component separation vs. frequency:}
It is interesting to note that for certain pulsars the 
cone associated with the emission seems to change
with frequency. For example, the conal emission in PSR 1920+21  
appears to have `switched' at 610 MHz to the innermost cone
while being associated with the central cone at other frequencies. 
Rankin (1983b), in a comprehensive
study of the dependence of component separation with 
frequency, invokes deep `absorption' features to explain the 
apparent anomalous reduction in the component separation
in certain frequency ranges.
We suggest that such anomalous reduction in the separations 
could be due to switching of the emission to an inner cone
at some frequencies. Observations at finely spaced frequencies
in the relevant ranges would 
be helpful to study this effect in detail. The other pulsars 
which show similar trends are PSRs 1804-08, 2003-08, 
1944+17 and 1831-04. It should be noted that such switching
is possibly reflected, also, in mode changes. 

{\bf The deficit at low $\beta/\rho_{\circ}$:} The absence 
of points near $\beta = 0$ is clearly noticeable in 
Fig.~\ref{fig:fig4}. Such a `gap' is also apparent in the 
distribution of $\beta/\rho_{\circ}$ plotted in
Fig.~\ref{fig:fig5}. 
The gap was already noted by 
Lyne \& Manchester (1988). They argued that it
arises because the rapid position-angle
swings (expected at small $\beta$'s) are difficult 
to resolve due to intrinsic or instrumental smearing, leading
to underestimation of the sweep-rates. With the improved quality
of data now available, the intrinsic smearing is likely to be
the dominant cause for this circumstance. 
There are a number of clear instances among the
general population of pulsars where the polarization angle traverse 
near the central core component is distorted. PSR 1237+25 
provides an extreme examples of such distortion, and 
Ramachandran \& Deshpande (1997) report promising initial 
efforts to model its polarization-angle track as distorted by 
by a low-$\gamma$ core-beam. 
Another possibility for the low-$\beta/\rho_{\circ}$ gap is that it
could simply be a selection effect caused by less intense emission
in the cone center than at intermediate traverses. If so, the 
low frequency turn-overs in the energy spectra of pulsars
may at least be partly due to this, since at lower radio 
frequencies the $\beta/\rho_{\circ}$ is relatively 
smaller.

{\bf The sources of uncertainties in the present analysis:} 
The standard deviation $\sigma_{\circ}$ corresponding 
to the best-fit model amounts to about 15\% of the  conal radius.
This fractional deviation (comparable to the thickness of the
cone) is too large to allow any more detailed description of the beam shape
(such as dependence on $\alpha$, for example). We find it 
useful to assess and quantify the sources of error, partly to
help possible refinement for future investigations. The 
three data inputs to our analysis are $\alpha, \beta$ and 
$\phi_{\nu}$, while the basic observables are the maximum
polarization-angle sweep rate and core width, apart from
the conal separation measured. It is easy to see that the 
errors in the core-widths will affect directly both 
$\alpha$ and $\beta$ estimates. Over the range of $\theta$s
spanned by the present data set the errors in $\alpha$ are 
likely to dominate, since the x \& y (in figure~\ref{fig:fig4})
are almost linearly proportional to $\sin(\alpha)$. Hence, the
fractional deviation may be nearly equal to 
(or define the upper limit of) the fractional error in $\sin(\alpha)$
and therefore in the core-width estimates.

Rankin (1990, 1993b) notes that in several cases the apparent 
core-widths might suffer from `absorption' and the widths might 
be underestimated if the effect is not properly accounted for.
Also, in some cases, the widths were extrapolated to a reference 
frequency of 1 GHz using a $\nu^{-0.25}$ dependence. There have
been several suggestions regarding the `appropriate' frequency 
dependence which would give significantly different answers 
when used for width extrapolation. For example, if our best-fit
dependence for conal width is used for the core-width extrapolation,
the values would differ from Rankin's estimates (through extrapolation)
by as much as
15\%, enough to explain the present deviation in some cases. 
Another possible 
source of error is the uncertainty in the sign of $\beta$ 
(important only for the $\sin(\alpha + \beta)$ term in 
equation~\ref{eq:mod} and hence for small $\alpha$).
As Rankin points out, it is difficult to determine the sign
unambiguously in most cases and hence the information is only 
available for a handful of pulsars.  

{\bf Evidence in favour of `conal' emission:} The significant
implication of the gap at $\theta\gsim 60^{\circ}$ (referred to
earlier) deserves further discussion. If the `conal' components
were results of a merely patchy (random) illumination across the
beam area, (as argued by Lyne \& Manchester, 1988), then such
a gap should not exist. 
If a single thick hollow 
cone were to be responsible for the conal components, a gap (corresponding
to the conal-single types) would still be apparent 
but then it should be above a cut-off y value (refer figure 4) 
and not in a angular sector like that observed.
On the other hand, if indeed the conal emission exists in the 
form of nested cones (as distinct from the core emission), 
then the shape of the gap is a natural consequence of our not including 
conal-single profiles in this analysis. This gap, therefore, should
be treated as an important evidence for a pulsar beam form comprised,
in general, of nested cones of emission.

\section{Summary}
Using the multifrequency pulse profiles of a large
number of conal-triple and multiple pulsars we  
modelled the pulsar beam shape in an improved way. 
Our analysis benefits from the
different frequency measurements being treated 
as independent samples, thus increasing the number of independent
constrains. The main results are summarized below.

1) Our profile sample is consistent with a beam shape that is 
a function of $\alpha$, circular at $\alpha =0$ and  
increasingly compressed in the latitudinal direction as
${\alpha}$ increases, as suggested by Biggs (1990). 
However, the data is equally 
consistent with the possibility that the beam is circular 
for all values of ${\alpha}$.

2) We identify three nested cones of emission based on a normalized
distribution of outer components. The observed gap
($\theta\gsim 60^{\circ}$) in the distribution 
independently suggests three cones in the form of annular rings whose 
widths are typically about 20\% of the cone radii.
We consider this circumstance as an important evidence for 
the nested-cone structure.

Any further significant progress in such modelling would necessarily 
need refined 
estimates of the observables, particularly the core-widths.

\section*{Acknowledgement} We thank V. Radhakrishnan, Rajaram 
Nityananda and Joanna Rankin for fruitful discussions and for several
suggestions that have helped in improving the manuscript.
We acknowledge Ashish Asgekar, D. Bhattacharya and  R. Ramachandran 
for useful discussions and thank our referee, J. A. Gil, for critical
comments and suggestions.

\end{document}